\documentclass[twocolumn,pra,aps,superscriptaddress,showpacs]{revtex4-1}

\usepackage{subfigure}
\usepackage{dcolumn}
\usepackage{amsmath,amssymb}
\usepackage{bm}
\usepackage{color}
\usepackage{latexsym}
\usepackage{epstopdf}
\usepackage{color}
\usepackage[english]{babel}
\usepackage{latexsym}
\usepackage{stmaryrd}

\usepackage{psfrag,graphicx} 
\usepackage{epsf} 
\usepackage{subfigure} 
\usepackage{amsmath} 
\usepackage{amssymb} 
\usepackage{amsfonts}
\usepackage{bm}
\usepackage{natbib}
\usepackage{epstopdf}
\usepackage{appendix}

\definecolor{mygrey}{gray}{0.}
\definecolor{myblue}{rgb}{0.2,0.2,0.8}
\definecolor{myzard}{cmyk}{0,0,0.05,0}
\definecolor{mywhite}{rgb}{1,1,1}
\definecolor{myred}{rgb}{1,0.,0.3}

\usepackage[colorlinks=true,citecolor=myblue,linkcolor=myred]{hyperref}
\def\be{\begin{equation}}
\def\ee{\end{equation}}
\def\ba{\begin{align}}
\def\enda{\end{align}}
\def\bi{\begin{itemize}}
\def\ei{\end{itemize}}

 \def\ee{\mathord{\rm e}}

 \def\ee{\mathord{\rm e}}

\def \be{\begin{equation}}
\def \ee{\end{equation}}
\def \ba{\begin{array}}
\def \ea{\end{array}}
\def \bea{\begin{eqnarray}}
\def \eea{\end{eqnarray}}

\newcommand{\ket}[1]{\left\vert #1 \right\rangle}
\newcommand{\bra}[1]{\left\langle #1 \right\vert}

\newcommand{\bla}[1]{\left( #1 \right)}
\newcommand{\blb}[1]{\left[ #1 \right]}

\begin{document}

\title{Long-lived driven solid-state quantum memory}

\author{J.-M. Cai}
\affiliation{Institut f\"{u}r Theoretische Physik, Albert-Einstein Allee 11, Universit\"{a}t Ulm, 89069 Ulm, Germany}
\affiliation{Center for Integrated Quantum Science and Technology, Universit\"{a}t Ulm, 89069 Ulm, Germany}
\author{F. Jelezko}
\affiliation{Center for Integrated Quantum Science and Technology, Universit\"{a}t Ulm, 89069 Ulm, Germany}
\affiliation{Institut f\"{u}r Quantenoptik, Albert-Einstein Allee 11, Universit\"{a}t Ulm, 89069 Ulm, Germany}
\author{N. Katz}
\affiliation{Racah Institute of Physics, The Hebrew University of Jerusalem, Jerusalem 91904, Givat Ram, Israel}
\author{A. Retzker}
\affiliation{Institut f\"{u}r Theoretische Physik, Albert-Einstein Allee 11, Universit\"{a}t Ulm, 89069 Ulm, Germany}
\affiliation{Racah Institute of Physics, The Hebrew University of Jerusalem, Jerusalem 91904, Givat Ram, Israel}
\author{M. B. Plenio}
\affiliation{Institut f\"{u}r Theoretische Physik, Albert-Einstein Allee 11, Universit\"{a}t Ulm, 89069 Ulm, Germany}
\affiliation{Center for Integrated Quantum Science and Technology, Universit\"{a}t Ulm, 89069 Ulm, Germany}

\date{\today}

\begin{abstract}
We investigate the performance of inhomogeneously broadened spin ensembles as quantum memories under continuous dynamical decoupling. The role of the continuous driving field is two-fold:  first, it decouples individual spins from magnetic noise; second and more important, it suppresses and reshapes the spectral inhomogeneity of spin ensembles. We show that a continuous driving field, which itself may also be inhomogeneous over the ensemble, can enhance the decay of the tails of the inhomogeneous broadening distribution considerably. This fact enables a spin ensemble based quantum memory to exploit the effect of cavity protection and achieve a much longer storage time. In particular, for a spin ensemble with a Lorentzian spectral distribution, our calculations demonstrate that continuous dynamical decoupling has the potential to improve its storage time by orders of magnitude for the state-of-art experimental parameters.
\end{abstract}

\maketitle

\section{Introduction}

Ensembles of a large number of emitters, e.g. cold atoms, polar molecules, and electronic spins, are very appealing systems for quantum memories \cite{QuanMen} and quantum repeaters \cite{QuanRep} for long distance quantum communication. Recently, experimental progress has led to the demonstration that hybrid quantum systems composed of superconducting qubits and a collection of electronic spins represents a promising new route \cite{Schu10,Kubo10,Wu10,Ams11,Zhu11}, which takes the advantage of long coherence times in spin ensembles and fast operation of a superconducting qubit. The large number, N, of spins provide a collective enhancement of the interaction strength between the spin ensemble and the electromagnetic field, and the collective coupling strength scales as $\Omega=(\sum_k g_k^2)^{1/2} \sim \sqrt{N} g$. The strong coupling regime for an ensemble of electronic spins and a superconducting resonator has been experimentally demonstrated recently\cite{Schu10,Kubo10,Wu10,Ams11}, i.e. $\sqrt{N} g \gg \gamma, \kappa$, where $\gamma$ and $\kappa$ are the spin's and resonator's damping rates.

Despite these achievements, the realisation of a quantum memory for microwave photons using spin ensembles remains a challenge due to both, the linewidth of the resonator and of the spin ensemble \cite{Kubo10}. Apart from the dephasing of individual spins due to magnetic noise, the spectral inhomogeneity over the ensemble will lead to the decay of the collective superradiant spin wave mode of the spin ensemble to dark states. The performance of a quantum memory is largely dependent on the properties of the spectral inhomogeneity \cite{Kur11,Din11}. In particular, if the tails of the spectral distribution decay faster than a Lorentzian profile, the cavity can provide coherence protection and improve the storage time of a quantum memory due to the collective enhancement of the coupling strength. However, Lorentzian distributions are found in many solid-state systems, e.g. electronic spin ensembles in diamond and impurities in crystals. In these systems, the memory time will still be limited significantly due to the spectral inhomogeneity and would hardly benefit from the strong collective coupling \cite{Kur11,Din11}.

\begin{figure}[t]
%\begin{minipage}{9cm}
%\hspace{-0.5cm}
\includegraphics[width=9cm]{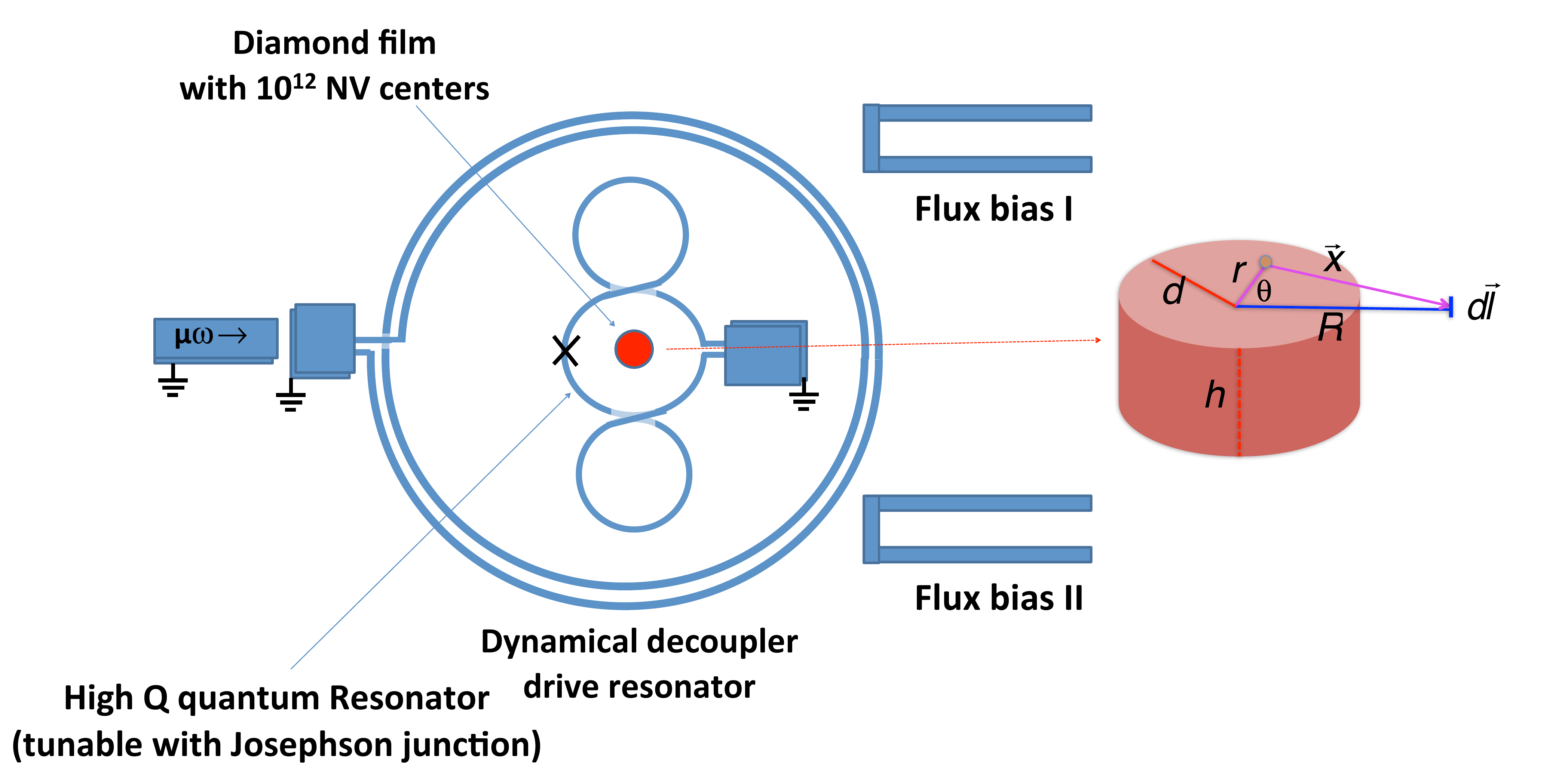}
%\includegraphics[width=3cm]{setup_b.pdf}
%\end{minipage}
\caption{(Color online) Setup of a cylinder-like diamond sample coupled with a high-Q quantum resonator. The dynamical decoupler drive resonator generates continuous microwave driving field. The driving field at the position with a distance $r$ to the center of the sample is calculated using the Biot-Savart law. The radius and height of the sample are denoted as $d$ and $h$ respectively. The distance from the center of the sample to the drive resonator is $R$.}\label{setup}
\end{figure}

In this work, we investigate how continuous dynamical decoupling can be used to construct a memory based on driven spin ensembles that has a long storage time. It is known that continuous driving fields can decouple spins from magnetic noise \cite{Tim08, Ber11,Cai11,Tim11,Ber12}. How continuous dynamical decoupling can improve the performance of quantum memories has not yet been examined in detail. Here, we show that besides the protection against magnetic field noise, continuous driving can have another prominent role in the present context, namely it can suppress and reshape the spectral inhomogeneity of spin ensembles. For Lorentzian distributions, continuous driving leads to a much faster decay of the tails of the effective spectral distributions. This effect makes all the spin transition frequencies far from the energy gap induced by the strong collective coupling, and thus the driven spin-ensemble memory can benefit from the cavity protection \cite{Kur11,Din11}. To analyze the performance of the driving system as a quantum memory, we calculate the decay of the collective Rabi oscillation and the state fidelity after a certain storage time by solving the Schr\"{o}dinger equation with the Laplace transformation. Our numerical calculations demonstrate that the storage time of a driven spin ensemble can be improved by orders of magnitude with reasonable experimental parameters, even if the continuous driving field itself is inhomogeneous over a finite size of sample.

\section{Setup of driven spin ensembles}

We consider a cylinder-shaped diamond sample, which is coupled to a superconducting resonator, see Fig.\ref{setup}. The radius and height of the sample are denoted as $d$ and $h$ respectively. The amplitude of the microwave driving field at the point with the distance $r$ to the center of the diamond sample, can be estimated using the Biot-Savart law $B (r)=\frac{g_e\mu_B\mu_0 I}{4\pi} \int\frac{\vec{dl}\times \vec{x}}{|\vec{x}|^3}$, and we get
\begin{equation}
B(r)=\frac{g_e\mu_B\mu_0 I}{4\pi} \int_0^{2\pi}\frac{R(R-r\cos\theta)}{(R^2-2Rr\cos\theta+r^2)^{3/2}} d\theta,
\end{equation}
which can be written as
\begin{equation} 
B(r)=\frac{g_e\mu_B \mu_0 I}{2R} \blb{1+\frac{3}{4}\bla{\frac{r}{R}}^2} + O[(\frac{r}{R})^4],
\label{eq:br}
\end{equation}
where $d\vec{l}$ is the differential element vector of the drive resonator in the direction of current, $\mu_0$ is the magnetic susceptibility, $R$ is the radius of the dynamical decoupler drive resonator, and $I$ is the amplitude of the current. It can be seen from Eq.(\ref{eq:br}) that when the condition $R >4 d$ is satisfied, the inhomogeneity of the driving field amplitude will be less than $5\%$ (i.e. the relative difference between the maximum and the minimum values across the sample). For diamond samples, it is feasible to achieve a  concentration of Nitrogen -Vacancy (NV) centers of $\rho_0=10^{15} \mbox{mm}^{-3}$ \cite{Aha09}. For a thin diamond film with the radius $d=0.1 \mbox{mm}$ and the depth $h=0.1\mbox{mm}$, the number of NV centers is around $10^{12}$. It has been estimated from experiments that the interaction between an individual spin and a superconducting resonator is of the order of $10 \mbox{Hz}$ \cite{Kubo10} and thus a collective coupling strength on the order of $10 \mbox{MHz}$ can be achieved. Assuming $\Delta$ the value of the width of the spectral distribution (which is of the order of $\mbox{MHz}$ \cite{Kubo10}), we can estimate that the required driving current is about $0.3 \bla{ \Delta/\mbox{MHz}} \mbox{A}$ in order to generate a microwave field (which creates a Rabi frequency) of $10 \Delta$ on the diamond sample with a driving source of radius $R=0.5 \mbox{mm}$. The required current can be reduced by using several resonator loops. These parameters are within reach of current experimental technology. Careful optimization may easily improve homogeneity and strengthen the drive. In the following, our calculations will use the conservative estimated driving parameters: $B_{min}=\frac{g_e\mu_B \mu_0 I}{2R}  =10 \Delta$ and $B_{max}=\frac{g_e\mu_B \mu_0 I}{2R} \blb{1+\frac{3}{4}\bla{\frac{d}{R}}^2}=B_{min}(1+0.05)$. We remark that the design of the $'8'$ shape high Q quantum resonator, see Fig.\ref{setup},  is to reduce the effect of the driving fields on the cavity mode since the the total driving field on the resonator should vanish.

\section{Effective spectral distribution of driven spin ensembles}

\begin{figure}[b]
\includegraphics[width=8cm]{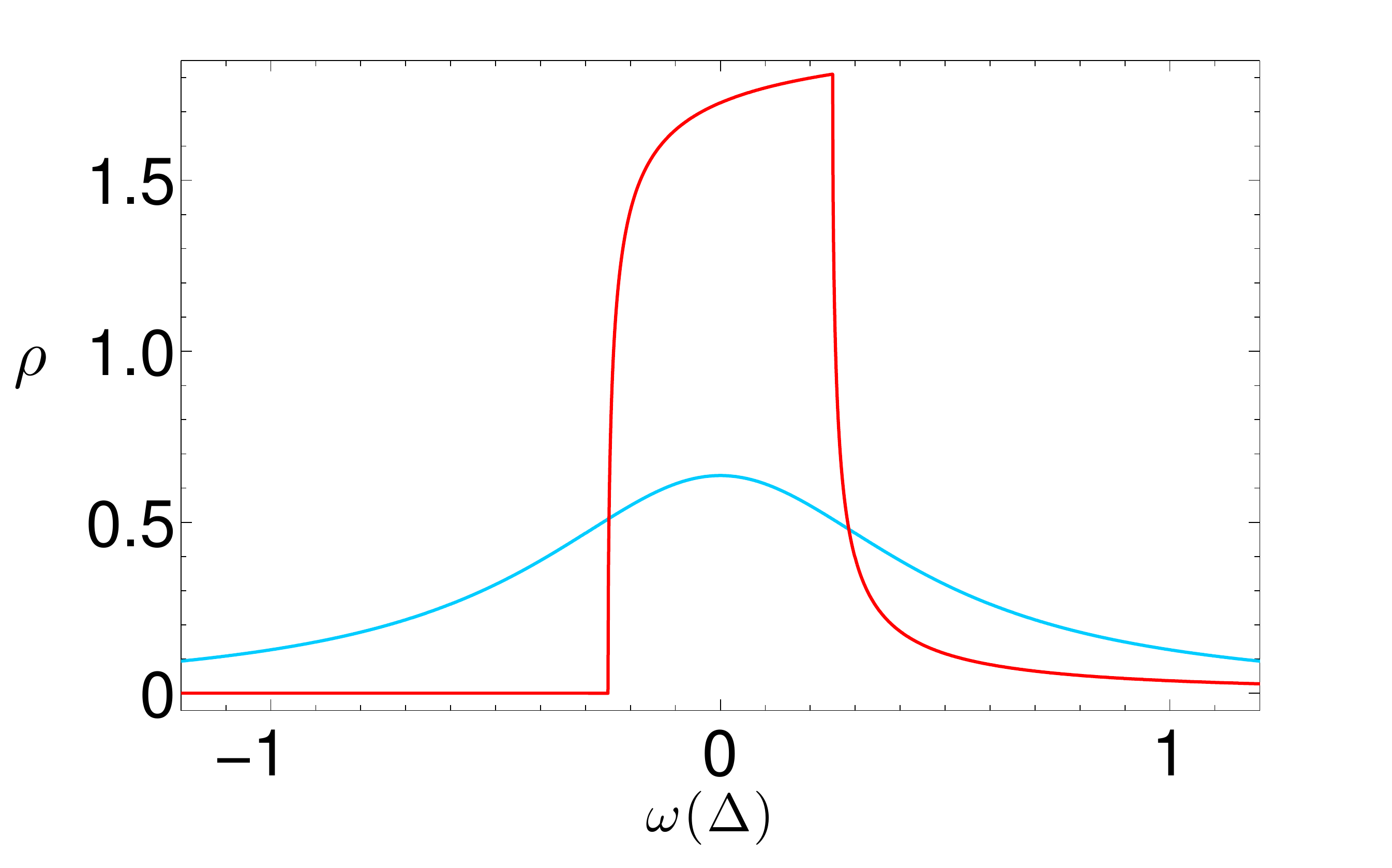}
\caption{Lorentzian spectral distribution (blue) and the one modified by continuous driving (red). We use the parameters $B_{min}=10 \Delta$ and $B_{max}=10.5 \Delta$. In the red curve the zero of frequency is located at $(B_{min}+B_{max})/2$.}\label{spectral}
\end{figure}

As we have discussed in the above section, the minimum and maximum values of the amplitude of the continuous driving field on the diamond sample are $B_{min}$ and $B_{max}$ respectively. We denote the probability function of the driving amplitude as $f(B)$, and have $f(B)dB=\rho_0 \cdot (2\pi r dr) h$ where $\rho_0$ is the concentration of NV center and $h$ is the depth of the diamond sample. We remark that the actual shape of the diamond sample will not affect the main conclusions of the following discussions. After some calculations, we can see that the distribution of $B$ actually has a rectangle profile, namely
\begin{equation}
f(B)dB =\rho_0 ( 2\pi r h) dr=\frac{4\pi(\rho_0 R^2 h)}{3B_{min}} dB.
\end{equation}
Thus, the normalized probability function for $B$ is $\bar{f} (B)=1/\Delta_B$ for $B\in [B_{min},B_{max}]$,  where $\Delta_B=B_{max}-B_{min}$ is the inhomogeneity of driving field. The local Hamiltonian of an individual electronic spin is
\begin{equation}
H_k=\frac{\omega_k}{2}\sigma_z^{(k)} + B_k \sigma_x^{(k)}\cos{(\omega_0 t)}
\end{equation}
where $\omega_0$ is chosen to coincide with the central value of the energy spectrum (which corresponds to the zero field splitting of NV center ground state namely $2.87 \mbox{GHz}$ and shifted by the applied magnetic field) and $\sigma_x^{(k)}, \sigma_z^{(k)}$ are Pauli operators. In the interaction picture, the effective Hamiltonian is written as
\begin{equation}
\bar{H}_k=\frac{\Delta_k}{2} \sigma_z^{(k)}+ \frac{B_k}{2} \sigma_x^{(k)}
\label{eq:Hk5}
\end{equation}
where we denote $\Delta_k=\omega_k-\omega_0$. After diagonalizing the above effective Hamiltonian, we obtain a dressed two-level system with
\begin{equation}
\bar{H}_k=\frac{\bar{\omega}_k}{2} (\ket{\uparrow}_k\bra{\uparrow}-\ket{\downarrow}_k\bra{\downarrow})
\end{equation}
with $\bar{\omega}_k=\sqrt{\Delta_k^2+B_k^2} \geq B_{min}$ where $\ket{\uparrow}_k$ and $\ket{\downarrow}_k$ denote the effective spin up and down states (namely the eigenstates of the effective Hamiltonian in Eq.(\ref{eq:Hk5})). The value of $B_k$ is uniformly distributed within the range $[B_{min},B_{max}]$, and we consider a Lorentzian distribution for $\Delta_k$, namely
\begin{equation}
\rho(\Delta_k)=\frac{2\Delta}{\pi\bla{\Delta^2+4\Delta_k^2}}.
\end{equation}
Thus, we can calculate the normalized probability distribution of the effective energy spectrum as follows
\begin{equation}
p(\bar{\omega}_k) d\bar{\omega}_k = 2 \int\limits_{B_{min}}^{\bar{\omega}_k} \bar{f}(B) \blb{ \rho\bla{\Delta_k=\sqrt{\bar{\omega}_k^2-B^2}}d\Delta_k} dB
\end{equation}
which leads to 
\begin{equation}
p(\bar{\omega}_k) d\bar{\omega}_k=\frac{4\Delta \bar{\omega}_k}{\pi} \int\limits_{B_{min}}^{\bar{\omega}_k} \bar{f}(B) \frac{\bla{\bar{\omega}_k^2-B^2}^{-1/2}}{\Delta^2+4\bla{\bar{\omega}_k^2-B^2}}dB d\bar{\omega}_k
\end{equation}
After performing the integration, we get
%\begin{widetext}
\begin{equation}
p(\bar{\omega}_k)  = \left\{ \begin{array}{lll}
         0 &   \bar{\omega}_k \in (-\infty, B_{min})\\
       \frac{4\bar{\omega}_k}{\pi \Delta_B} \frac{\frac{\pi}{2}-\mu(B_{min},\bar{\omega}_k)}{\sqrt{4\bar{\omega}_k^2+\Delta^2}}&    \bar{\omega}_k\in [B_{min},  B_{max}]\\
 \frac{4\bar{\omega}_k}{\pi\Delta_B} \frac{\mu(B_{max},\bar{\omega}_k)-\mu(B_{min},\bar{\omega}_k)}{\sqrt{4\bar{\omega}_k^2+\Delta^2}}&  \bar{\omega}_k \in (B_{max},\infty) 
\end{array} \right.
\end{equation}
%\end{widetext}
where $\mu(B,\bar{\omega}_k)=\tan^{-1} (B\Delta/\sqrt{(4\bar{\omega}_k^2+\Delta^2)(\bar{\omega}_k^2-B^2)})$ and $\Delta_B$ is the inhomogeneity of the driving field. In the ideal case of perfectly homogeneous driving, namely when $B_{max}=B_{min}$, the distribution function is $p(\bar{\omega}_k) =4\bar{\omega}_k\Delta /\pi \sqrt{\bar{\omega}_k^2-B_{min}^2} \blb{\Delta^2+4\bla{\bar{\omega}_k^2-B_{min}^2}} $ for $\bar{\omega}_k> B_{min}$.
The interaction between individual spins and the field for a resonant cavity is
\begin{equation}
H_{k-c}=g_k(a\sigma_+^k+a^{\dagger} \sigma_-^k)
\end{equation}
where $\sigma_{\pm}^{(k)}=(\sigma_x^{(k)}\pm i\sigma_y^{(k)})/2$ and $g_k \sim 10 \mbox{Hz}$. With the rotating wave approximation, which is valid when $g_k \ll X_0$, we can obtain the effective interaction $\bar{H}_{k-c}=\bar{g}_k(a\ket{\uparrow}_k\bra{\downarrow}+a^{\dagger} \ket{\downarrow}_k\bra{\uparrow})$ between the cavity field and the dressed two-level system with
\begin{equation}
\bar{g}_k=\frac{1}{2}g_k\bla{1+\frac{\Delta_k}{\bar{\omega}_k}}.
\end{equation}
Taking into account the distribution of $\Delta_k$, one can show that the correction of $\frac{\Delta_k}{\bar{\omega}_k}$ is negligible when $B_k \gg \Delta_k$ \cite{footnote1}, and thus the effective coupling strength can be approximated as $\bar{g}_k=g_k/2$ (i.e. $\bar{\Omega}=(\sum_k \bar{g}_k^2)^{1/2}=\Omega/2$).

\begin{figure}[b]
\begin{minipage}{9cm}
\hspace{-0.2cm}
\includegraphics[width=4.3cm]{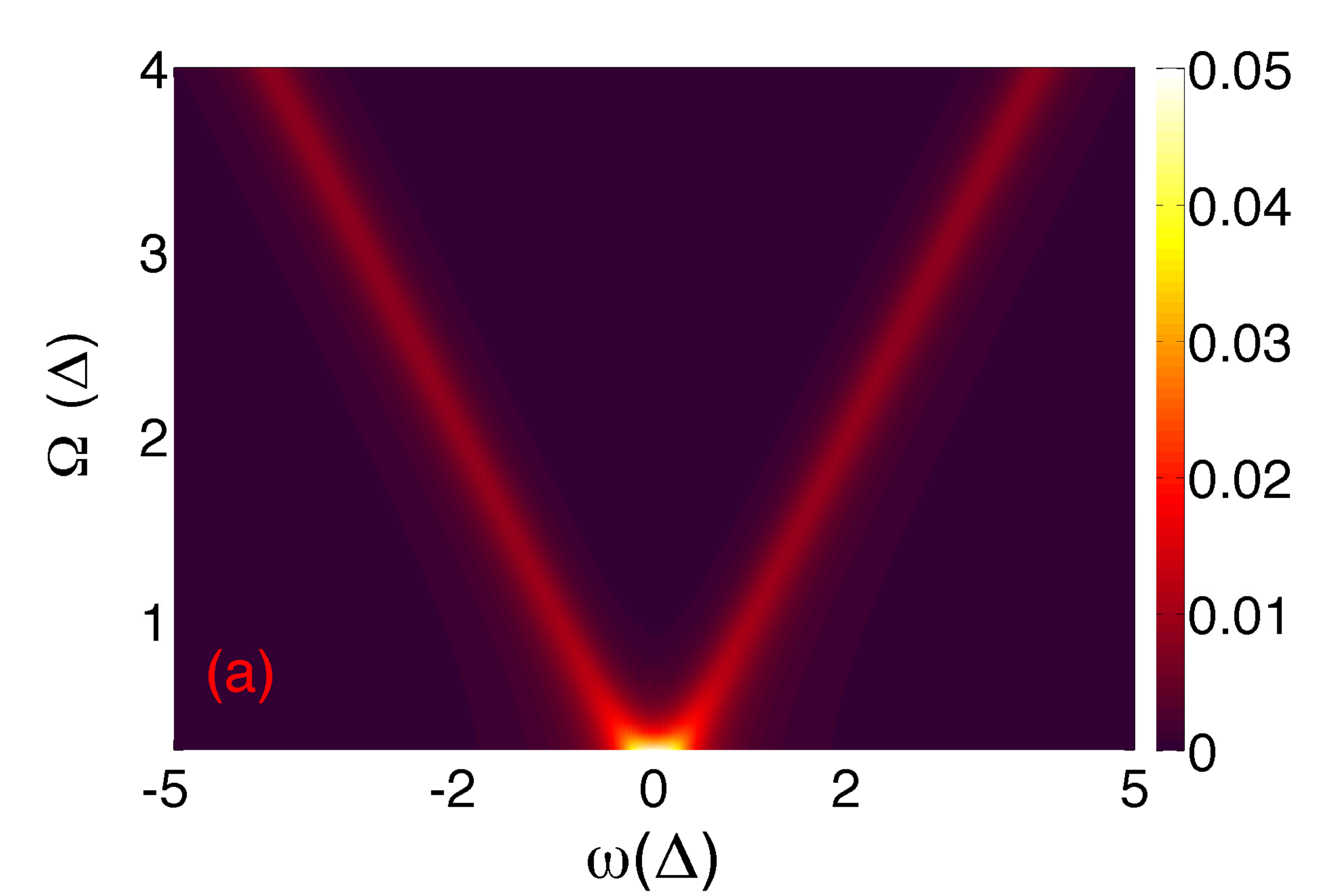}
\hspace{-0.3cm}
\includegraphics[width=4.3cm]{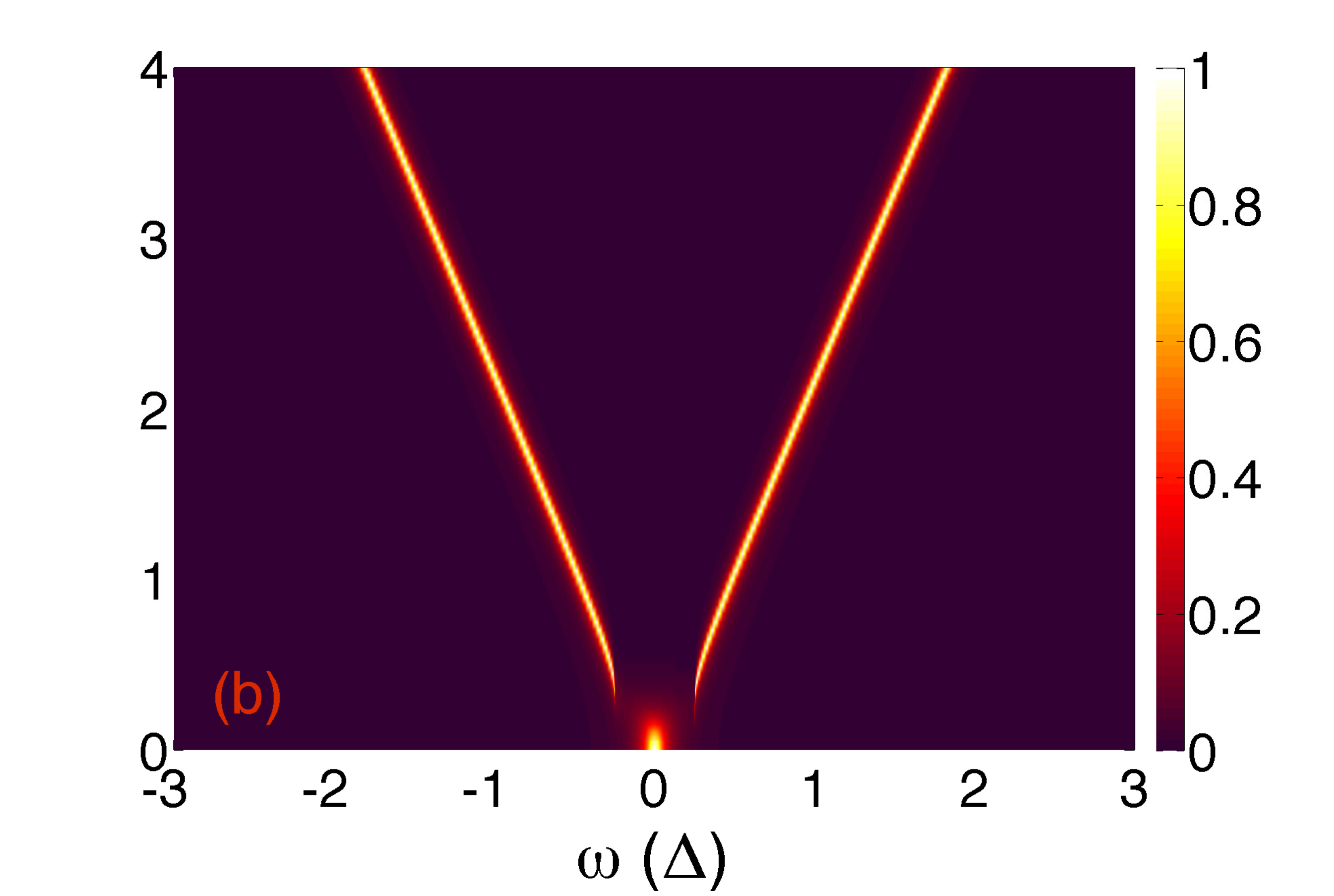}
\end{minipage}
\begin{minipage}{9cm}
\hspace{-0.6cm}
\includegraphics[width=8.5cm]{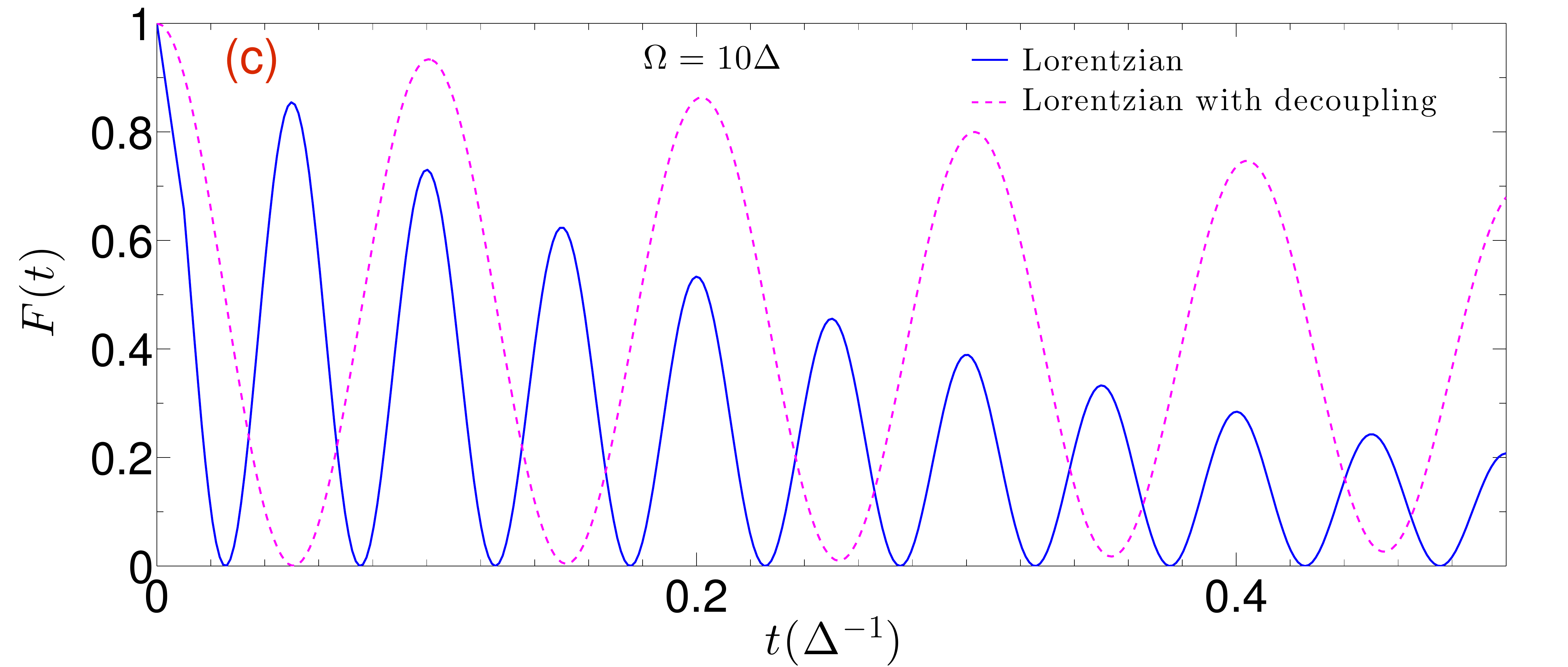}
\end{minipage}
\caption{Transmission of a resonator (as in Eq.(\ref{eq:tf})) coupled to a  Lorentzian spectral distribution of spins without (a) and with (b) continuous driving. The parameters are $B_{min}=10 \Delta$ and $B_{max}=10.5 \Delta$, and $\kappa=0.1\mbox{MHz}$, $\gamma=10^{-4} \mbox{MHz}$. (c)The collective Rabi oscillation between the cavity and the superradiant spin wave mode with the initial state $\ket{1,G}$. The curves show the quantity $F(t)=|\bra{1,G} \exp{(-it H)} \ket{1,G}|^2$ as a function of time.}\label{rabi}
\end{figure}

\section{Transmission function and Collective Rabi oscillation}

We consider an ensemble of electronic spins coupled to a resonator, whose frequency and damping rate are denoted $\omega_c$ and $\kappa$ respectively. When the number of excitations is low as compared to the total number of electronic spins, the spins can be described as a collection of oscillators in the Holstein-Primakoff approximation with $\sigma_z^{(k)}=a_k^{\dagger}a_k-\frac{1}{2}$ and $\sigma_+^{(k)}=a_k^{\dagger} \sqrt{1-a_k^{\dagger}a_k}\approx a_k^{\dagger} $, thus the total Hamiltonian can be written as
\begin{equation}
H=\omega_c a_c^{\dagger } a_c+ \sum_{k=1}^{N} \omega_k a_k ^{\dagger }a_k +\sum_{k=1}^{N} g_k (a_k^{\dagger } a_c + a_k a_c^{\dagger})
\label{eq:H1}
\end{equation}
Under the driving of continuous field, we have a collection of dressed two-level systems coupled with the resonator, and the effective Hamiltonian is
\begin{equation}
\bar{H}=\omega_c a_c^{\dagger } a_c+ \sum_{k=1}^{N} \bar{\omega}_k  \bar{a}_k ^{\dagger }\bar{a}_k+\sum_{k=1}^{N} \frac{g_k}{2} (\bar{a}_k^{\dagger } a_c +  \bar{a}_k a_c^{\dagger}  )
\label{eq:Hbar}
\end{equation}
The difference between Eq.(\ref{eq:H1}) and Eq.(\ref{eq:Hbar}) is that the original Lorentzian distribution of $\omega_k$ is now modified to the new distribution of $\bar{\omega}_k$, as shown Fig.\ref{spectral}. The width of the distribution depends on the inhomogeneity of the driving field, and most important the decay of its tail becomes much faster. This fact will greatly improve the performance of such a quantum memory as we will show in the rest of this work, although the effective coupling is decreased by half.

We assume that the cavity is on resonance with the spins, namely its frequency is $\omega_c=\omega_0$ for the case with no continuous driving; while the cavity resonant frequency is $\omega_c=\omega_0+(B_{min}+B_{max})/2$ corresponding to the dressed spin wave mode under continuous driving. The transmission function of the cavity from the standard input-out theory is given by \cite{Din11,Tuf12}
\begin{equation}
t(\omega)=\frac{i\kappa}{2}\frac{1}{\bla{\omega_0-i\kappa/2-\omega}-\sum\limits_k g_k^2 / \bla{\omega_k-i\gamma/2-\omega}}
\label{eq:tf}
\end{equation}
where $\gamma$ is the damping rate of spins. In the strong coupling regime, namely the collective coupling is much larger than the decay rates (i.e. $\Omega \gg \gamma,\kappa$), the excitation is coherently exchanged between the cavity and the superradiant mode of the ensemble of emitters. The so-called vacuum Rabi splitting corresponds to the appearance of double peaks in the transmission spectrum centred at the frequencies around $\pm \Omega$. Inhomogeneous broadening will damp the collective Rabi oscillation and lead to the line broadening of the transmission peaks. In Fig.\ref{rabi}(a-b), we show that the linewidth of the transmission peaks is greatly narrowed with continuous driving field, because the inhomogeneous broadening is suppressed. Once the effective collective coupling $\bar{\Omega}$ is larger than the inhomogeneity of the driving field $B_{max}-B_{min}$, the transmission function shows two peaks, the linewidth of which is mainly determined by the damping rate of the cavity and individual spins. Such a behaviour is very similar to the ideal rectangular spectral distribution \cite{Din11}. To compare the effect of inhomogeneous broadening on the collective Rabi oscillation, we assume that the damping rate $\kappa=0$ and $\gamma=0$. In Fig.\ref{rabi}(c), it can be seen that the coherence time of the collective Rabi oscillation is greatly improved with continuous driving.

\section{Long-lived quantum memory}

The strong coupling between the cavity and the ensemble induces an energy gap between the dressed polariton modes, i.e. superpositions of the cavity mode $a_c$ and the ensemble superradiant mode $b=\sum_k (g_k/\Omega) a_k$, which are protected from the dephasing caused by the inhomogeneous broadening. With such an effect of cavity protection, it is possible to build a quantum memory with a storage time much longer than the lifetime of the cavity mode and the dephasing time of the spin ensemble. The condition for the cavity protection is that the spectral distribution should decay sufficiently fast so that all the spin transition frequencies are far from the energy gap (arising from the collective coupling) \cite{Kur11,Din11}. This explains why a Lorentzian distribution of spin ensemble can not benefit much from the cavity protection, in particular as compared to the rectangular spectrum profile.

\begin{figure}[t]
\begin{minipage}{9cm}
\hspace{-0.5cm}
\includegraphics[width=4.6cm]{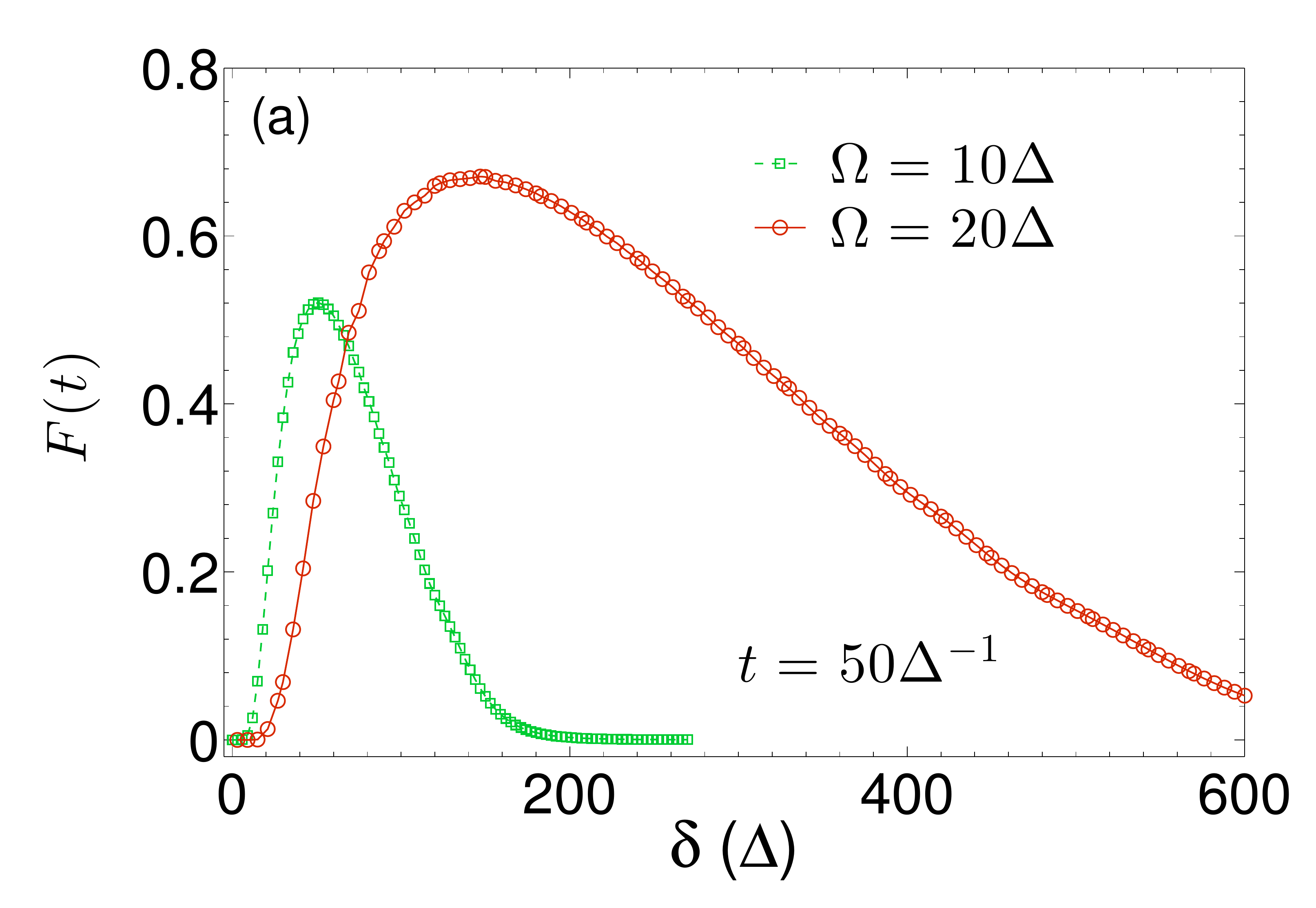}
\hspace{-0.3cm}
\includegraphics[width=4.6cm]{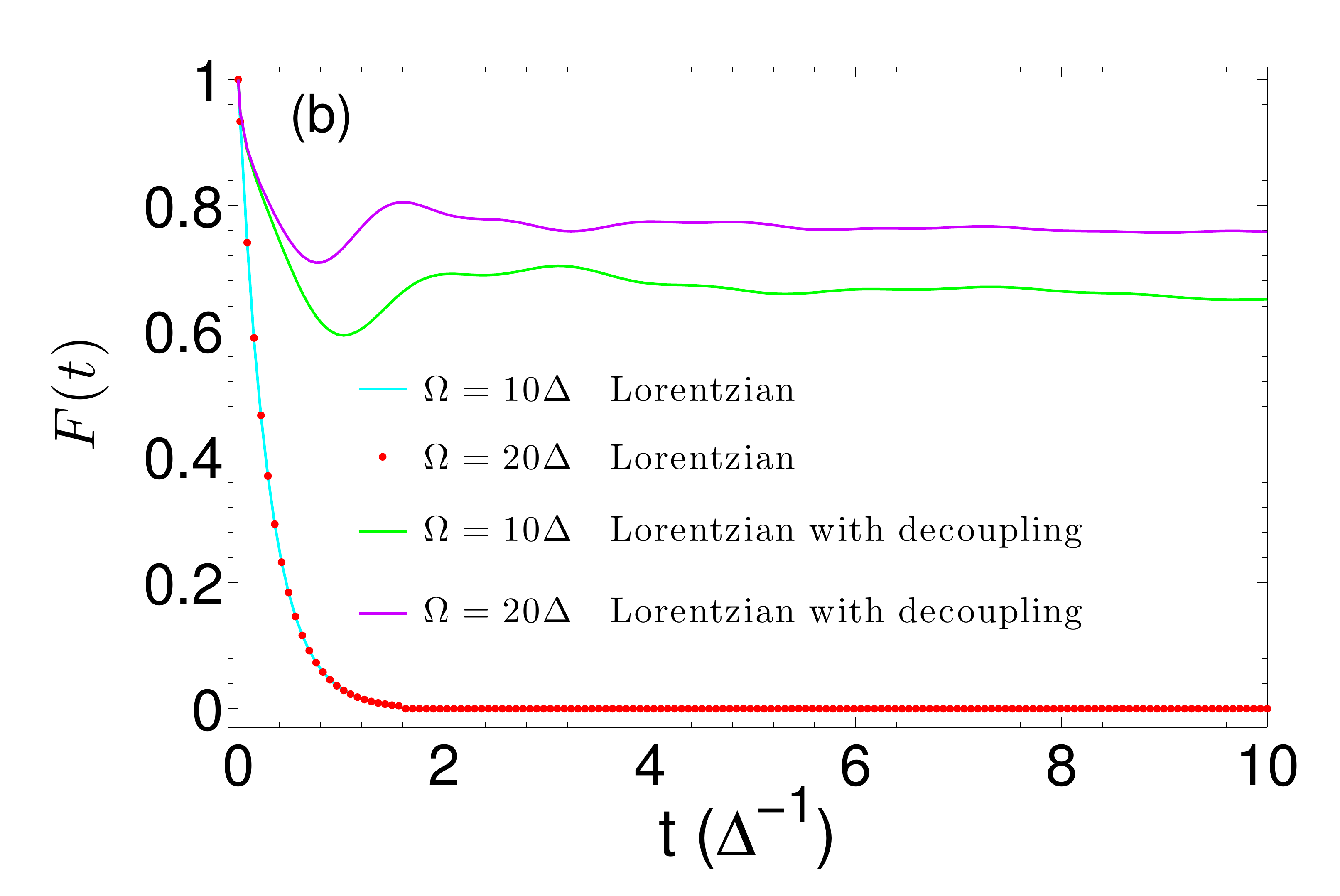}
\end{minipage}
\caption{(a) The state fidelity $F$ at time $t=50 \Delta ^{-1}$ as a function of the cavity-spin detuning $\delta$. (b) The time evolution of the state fidelity $F(t)$ for a Lorentzian quantum memory without and with continuous driving. The detuning $\delta$ is chosen as the optimal value.}\label{mem}
\end{figure}

The other critical factor for a long-lived quantum memory is the damping of the cavity. To overcome this problem, one can choose the cavity-spin tuning $\delta$ such that the effect of cavity damping is suppressed but the memory can still benefit from the cavity protection \cite{Kur11,Din11}. The microwave photon can be stored to the dressed polariton mode as
\begin{equation}
\ket{\psi_0(\delta)}=\cos\bla{\frac{\theta}{2}}\ket{1,G}-\sin\bla{\frac{\theta}{2}}\ket{0,S}
\end{equation}
where $\cot\bla{\theta}=(\delta/2 \Omega)$ and $\ket{S}=(1/\Omega) \sum_k g_k \ket{\uparrow_k}$. Thus, if the detuning is large the excitation is more likely to stay in the spin superradiant mode and suffers less from the damping of the cavity. While the cavity protection against the inhomogeneous broadening requires a small detuning so that the effective collective coupling is strong. The storage time of a quantum memory achieves the best value for an optimal detuning \cite{Din11}. The performance of a quantum memory is characterized by the state fidelity $F(t)=|f(t)|^2$ with $f(t)=\bra{\phi_0 (\delta)}\phi (\delta,t)\rangle$,  where the state at time $t$ is $\ket{\phi (\delta,t)}=\exp{(-it H)} \ket{\phi_0 (\delta)}=\alpha_0 \ket{1,G}+\sum_{k=1}^{N}\alpha_k \ket{0,\uparrow_k}$. To calculate the state fidelity, one needs to solve the Schr\"{o}dinger equation \cite{Kur11,Din11} for the coefficients $\alpha_k$, namely
\begin{eqnarray}
\dot{\alpha_0} (t)&=&-i\tilde{\omega}_c \alpha_0 (t) -i\sum_{k=1}^N g_k \alpha_k(t)\\
\dot{\alpha_k} (t)&=& -i g_k \alpha_0(t) -i\tilde{\omega}_k \alpha_k (t).
\end{eqnarray}
where $\tilde{\omega}_c =\omega_c-i\kappa/2$, $\tilde{\omega}_k=\omega_k-i\gamma/2$. By performing the Laplace transformation, we obtain
\begin{eqnarray}
s A_0(s)-A_0(0)&=&-i\tilde{\omega}_c A_0(s) -i\sum_{k=1}^N g_k A_k(s)\\
s A_k (s)-A_k(0)&=& -i g_k A_0(s) -i\tilde{\omega}_k A_k (s).
\end{eqnarray}
The solutions of $A_0(s)$ and $A_k(s)$ are
\begin{eqnarray}
A_0(s)&=&\blb{\cos(\frac{\theta}{2})+\frac{i}{\Omega} \sin(\frac{\theta}{2})  M(s)} T(s)\\
A_k(s)&=& - \frac{i g_k}{s+i\tilde{\omega}_k} A_0(s) -\frac{1}{\Omega} \sin(\frac{\theta}{2})  \frac{g_k}{s+i\tilde{\omega}_k}.
\end{eqnarray}
where
\begin{equation}
M(s)=\sum_{k=1}^{N} \frac{g_k^2}{s+i\tilde{\omega}_j}, \quad \mbox{and} \quad T(s)=\frac{1}{s+i\tilde{\omega}_c+M(s)}.
\end{equation}
The Laplace transformation of $f(t)=\bra{\phi_0 (\delta)}\phi (\delta,t)\rangle=\cos(\frac{\theta}{2}) \alpha_0(t)-(1/\Omega)\sin(\frac{\theta}{2}) \sum_k g_k \alpha_k(t)$ is
\begin{equation}
\bar{f}(s)=\cos(\frac{\theta}{2}) A_0(s)-(1/\Omega)\sin(\frac{\theta}{2}) \sum_{k=1}^{N} g_k A_k(s)
\label{eq:fts}
\end{equation}
from which one can calculate the state fidelity $F(t)$ and compare the storage time of a quantum memory without and with continuous driving.

We have compared the state fidelity for a Lorentzian distribution of spin ensemble with and without continuous driving, namely calculated the solution in Eq.(\ref{eq:fts}) for the original Hamiltonian in Eq.(\ref{eq:H1}) and the dressed Hamiltonian in Eq.(\ref{eq:Hbar}). It can be seen from Fig.\ref{mem}(a-b) that: (1) The increasing collective coupling has very little effect on the state fidelity for Lorentzian distribution if there is no driving as expected;  (2) The storage time can be improved by orders of magnitude with continuous driving; (3) The continuous driving also enables the quantum memory to benefit from the cavity protection, which suppresses the effect of the inhomogeneity in the driving field and thereby improves the performance with increasing collective coupling strength. We stress that the parameters in our numerical simulations are within the state-of-art cavity QED technologies. Careful optimization may easily improve homogeneity and strengthen the drive. To be more specific, we conservatively take the inhomogeneous broadening as $\Delta=1\mbox{MHz}$ and the damping rate of the cavity $\kappa=0.1\mbox{MHz}$ (i.e. the cavity lifetime is $10 \mu s$). The storage time of a Lorentzian quantum memory is limited by the dephasing of the spin ensemble (about $1\mu s$). With continuous driving $B_{min}=10 (20)\mbox{MHz}$ and $B_{max}=10.5 (20.5) \mbox{MHz}$, the state fidelity can be as high as $0.5 (0.7)$ even after $50\mu s$.

\section{Summary and discussion}

In summary, we have shown that continuous driving can suppress the effect of inhomogeneous broadening on a quantum memory based on spin ensembles. Besides, the obvious role of decoupling individual spins from magnetic noise, continuous driving can also reduce and reshape the spectral distribution. Even if the continuous driving field itself is inhomogeneous over the ensemble, it can make the tails of the spectral density profile decay much faster and approach the ideal rectangle distribution. This feature then allows us to take  advantage of the cavity protection in the strong coupling regime and to achieve much improved performance of a quantum memory. We consider the Lorentzian spectral distribution as an example, as it is relevant in many solid-state systems, e.g. spin ensemble in diamond. Our results demonstrate that the linewidth of the double peaks in the transmission spectrum is greatly narrowed by continuous driving. Furthermore, the storage time of a driven quantum memory can be extended by orders of magnitude as compared with the one without driving. We stress that the main feature of our scheme is that the effect of the inhomogeneity of continuous driving field can be suppressed by the strong collective coupling via cavity protection. As comparison, one could invert all the spins half way through the storage period to eliminate the inhomogeneous broadening, nevertheless, the pulse inhomogeneity will inevitably lead to inversion errors over individual spins which can not directly be compensated by cavity protection as occurs for continuous driving. We expect that the idea of driven quantum memory can be extended to other physical systems, e.g. ion Coulomb crystals \cite{Her09,Albert11}.

\section{Acknowledgements}  

We thank Igor Diniz for helpful discussions. The work was supported by the Alexander von Humboldt Foundation, the EU Integrating Project Q-ESSENCE, the EU STREP PICC and HIP, DIAMANT, the BMBF Verbundprojekt QuOReP, DFG (FOR 1482 and 1493), ISF 1248/10 and DARPA. J.-M.C was supported also by a Marie-Curie Intra-European Fellowship within the 7th European Community Framework Program.


\begin{thebibliography}{99}

\bibitem{QuanMen} C. Simon, M. Afzelius, J. Appel, A. Boyer de la Giroday, S.J. Dewhurst, N. Gisin, C.Y. Hu, F. Jelezko, S. Kroll, J.H. Muller, J. Nunn, E. Polzik, J. Rarity, H. de Riedmatten, W. Rosenfeld, A.J. Shields, N. Skold, R.M. Stevenson, R. Thew, I. Walmsley, M. Weber, H. Weinfurter, J. Wrachtrup, R.J. Young, \emph{Quantum memories: A review based on the European integrated project Qubit Applications (QAP)}, Eur. Phys. J. D {\bf 58}, 1 (2010), and the references therein.

\bibitem{QuanRep} N. Sangouard, C. Simon, H. de Riedmatten, and N. Gisin, \emph{Quantum repeaters based on atomic ensembles and linear optics}, Rev. Mod. Phys.  {\bf 83}, 33 (2011).

\bibitem{Schu10} D. I. Schuster, A. P. Sears, E. Ginossar, L. DiCarlo, L. Frunzio, J. J. L. Morton, H. Wu, G. A. D. Briggs, B. B. Buckley, D. D. Awschalom, and R. J. Schoelkopf, \emph{High-Cooperativity Coupling of Electron-Spin Ensembles to Superconducting Cavities}, Phys. Rev. Lett.  {\bf 105}, 140501 (2010).

\bibitem{Kubo10} Y. Kubo, F. R. Ong, P. Bertet, D. Vion, V. Jacques, D. Zheng, A. Dr\'{e}au, J.-F. Roch, A. Auffeves, F. Jelezko, J. Wrachtrup, M. F. Barthe, P. Bergonzo, and D. Esteve, \emph{Strong Coupling of a Spin Ensemble to a Superconducting Resonator}, Phys. Rev. Lett.  {\bf  105}, 140502 (2010).

\bibitem{Wu10} H. Wu, R. E. George, J. H. Wesenberg, K. M\o lmer, D. I. Schuster, R. J. Schoelkopf, K. M. Itoh, A. Ardavan, J. J. L. Morton, and G. A. D. Briggs, \emph{Storage of Multiple Coherent Microwave Excitations in an Electron Spin Ensemble}, Phys. Rev. Lett.  {\bf 105}, 140503 (2010).

\bibitem{Ams11} R. Ams\"{u}ss, Ch. Koller, T. N\"{o}bauer, S. Putz, S. Rotter, K. Sandner, S. Schneider, M. Schramb\"{o}ck, G. Steinhauser, H. Ritsch, J. Schmiedmayer, and J. Majer, \emph{Cavity QED with Magnetically Coupled Collective Spin States}, Phys. Rev. Lett.  {\bf  107}, 060502 (2011).

\bibitem{Zhu11} X. Zhu, S. Saito,	A. Kemp, K. Kakuyanagi, S. Karimoto,	H. Nakano, W. J. Munro,	Y. Tokura,	M. S. Everitt, K. Nemoto, M. Kasu, N. Mizuochi and K. Semba, \emph{Coherent coupling of a superconducting flux qubit to an electron spin ensemble in diamond}, Nature {\bf 478}, 221 (2011).

\bibitem{Kur11} Z. Kurucz, J. H. Wesenberg, and K. M\o lmer, \emph{Spectroscopic properties of inhomogeneously broadened spin ensembles in a cavity}, Phys. Rev. A {\bf 83}, 053852 (2011).

\bibitem{Din11} I. Diniz, S. Portolan, R. Ferreira, J. M. G\'{e}rard, P. Bertet, and A. Auff$\grave{e}$ves, \emph{Strongly coupling a cavity to inhomogeneous ensembles of emitters: Potential for long-lived solid-state quantum memories}, Phys. Rev. A {\bf 84}, 063810 (2011).

\bibitem{Tim08} N. Timoney, V. Elman, S. Glaser, C. Weiss, M. Johanning, W. Neuhauser, and Chr. Wunderlich, \emph{Error-resistant single-qubit gates with trapped ions}, Phys. Rev. A {\bf 77}, 052334 (2008).

\bibitem{Ber11} A. Bermudez, F. Jelezko, M. B. Plenio, A. Retzker, \emph{ Electron-Mediated Nuclear-Spin Interactions Between Distant NV Centers }, Phys. Rev. Lett. {\bf 107}, 150503 (2011).

\bibitem{Ber12} A. Bermudez, P. O. Schmidt, M. B. Plenio, A. Retzker, \emph{ Robust Trapped-Ion Quantum Logic Gates by Continuous Dynamical Decoupling }, Phys. Rev. A {\bf 85}, 040302 (2012).

\bibitem{Cai11} J.-M. Cai, B. Naydenov, R. Pfeiffer, L. P. McGuinness, K. D. Jahnke, F. Jelezko, M. B. Plenio, A. Retzker, \emph{ Robust dynamical decoupling with concatenated continuous driving }, E-print arXiv:1111.0930.

\bibitem{Tim11} N. Timoney, I. Baumgart, M. Johanning, A. F. Varon, M. B. Plenio, A. Retzker, Ch. Wunderlich, \emph{Quantum gates and memory using microwave-dressed states}, Nature {\bf 476}, 185 (2011).

\bibitem{Aha09}  I. Aharonovich, C. Santori, B. A. Fairchild, J. Orwa, K. Ganesan, K.-M. C. Fu, R. G. Beausoleil, A. D. Greentree, and S. Prawer, \emph{Producing optimized ensembles of nitrogen-vacancy color centers for quantum information applications }, J. Appl. Phys. {\bf 106}, 124904 (2009).

\bibitem{footnote1} A spin ensemble can be described by the coupling-density profile as $\rho_g(\omega)\equiv\sum_k\bar{g}_k^2 \delta(\omega-\bar{\omega}_k)  =\sum_k (\frac{g_k}{2})^2 [1+2\frac{\Delta_k}{\bar{\omega}_k}+(\frac{\Delta_k}{\bar{\omega}_k})^2]\delta(\omega-\bar{\omega}_k)$. Note that the probability distribution of $\Delta_k=\omega_k-\omega_0$ is symmetric for positive and negative values of $\Delta_k$, therefore we have $\rho_g(\omega)\simeq \sum_k (\frac{g_k}{2})^2 [1+(\frac{\Delta_k}{\bar{\omega}_k})^2]\delta(\omega-\bar{\omega}_k)$, and the correction is thus the second-order of $(\frac{\Delta_k}{\bar{\omega}_k})$, which is negligible when the driving is much stronger than the original spectral broadening.

\bibitem{Tuf12} T. Tufarelli, A. Retzker, M.B. Plenio, A. Serafini, \emph{ Input-output Gaussian channels: theory and application}, E-print arXiv:1205.6004.

\bibitem{Her09} P. F. Herskind, A. Dantan, J. P. Marler, M. Albert and M. Drewsen, \emph{Realization of collective strong coupling with ion Coulomb crystals in an optical cavity}, Nature Physics {\bf 5}, 494 (2009).

\bibitem{Albert11} M. Albert, J. P. Marler, P. F. Herskind, A. Dantan, and M. Drewsen, \emph{Collective strong coupling between ion Coulomb crystals and an optical cavity field: Theory and experiment}, Phys. Rev. A {\bf 85}, 023818 (2012).



\end{thebibliography}
\end{document}